\documentstyle[12pt]{article}

\def\lapproxeq{\lower .7ex\hbox{$\;\stackrel{\textstyle <}{\sim}\;$}}
\def\gapproxeq{\lower .7ex\hbox{$\;\stackrel{\textstyle >}{\sim}\;$}}

\begin{document}

\titlepage

\begin{flushright}
DTP/94/116 \\ December 1994
\end{flushright}
\vspace*{2cm}

\begin{center}
\vspace*{2cm}
{\Large{\bf Coulomb effects in $W^+W^-$ production}}
\end{center}

\vspace*{1cm}
\begin{center}
V.S.\ Fadin$^\dagger$, V.A.\ Khoze and A.D.\ Martin\\
\vspace*{0.3cm}
Department of Physics, University of Durham, Durham, DH1 3LE, U.K.\\
\vspace*{0.3cm}
and\\
\vspace*{0.3cm}
A.\ Chapovsky\\
\vspace*{0.3cm}
Novosibirsk State University, 630090 Novosibirsk, Russia.\\
\end{center}

\vspace*{2cm}

\begin{abstract}
We calculate the Coulomb effects on the cross section for
$e^+ e^- \rightarrow W^+ W^-$ taking into account the instability of the
$W$ bosons. We carefully explain the consequences of instability throughout the
energy range which will be accessible at LEP2. We present a formula which
allows
these effects to be easily implemented.
\end{abstract}

\vfil
\noindent ---------------\hfil

\noindent{\footnotesize{$^\dagger$ Permanent address: Budker Institute for
Nuclear Physics and Novosibirsk State University, 630090 Novosibirsk,
Russia.}}
\newpage

\noindent {\large {\bf 1.  Introduction}}
\vspace*{.5cm}

The detailed measurements of the observables associated with the $Z$ boson at
LEP and SLC, together with the known values of the QED and Fermi couplings
($\alpha$, $G_F$), have allowed precision tests of the electroweak
sector of the Standard Model. If this information were supplemented by an
accurate measurement of the mass of the $W$ boson, $M_W$, then extremely
stringent tests of the model would be possible. Much progress has been made
in the measurement of the mass $M_W$ from the study of $W \rightarrow \ell \nu$
events at the Fermilab $p \bar p$ collider \cite{CDF}, although the precision
is considerably less
than that for $\alpha$, $G_F$ and $M_Z$. One of the main objectives of the
LEP2 physics programme is to improve our knowledge of $M_W$.

The methods which have been proposed to measure $M_W$ from observations of
the process $e^+ e^- \rightarrow W^+ W^-$ at LEP2 include (i) the
direct reconstruction of $M_W$ from the decay products of the $W$ bosons, (ii)
the study of the end-point of the lepton spectrum in $W \rightarrow \ell \nu$
decays and (iii) an energy scan of the cross section in the $W^+ W^-$
threshold region. Methods (i) and (iii) are expected to be most
precise, although both have advantages and disadvantages. For method (i)
LEP2 can be run at its highest proposed energy to maximize the event rate,
but to reconstruct
the $W^+ W^- \rightarrow q_1 \bar q_2 q_3 \bar q_4$ decay channels we
encounter the problem of attributing all the observed decay products to the
correct parent $W$, and thus we need to control the QCD interference
(interconnection) effects \cite{SK}. These problems are reduced for the
$W^+ W^- \rightarrow q_1 \bar q_2\ell \nu$ decay channels, but then the
event rate is smaller and moreover an unobservable neutrino is present.
The threshold
scan, method (iii), is theoretically cleaner, but in the important region,
within about a $W$ width of the $W^+ W^-$ threshold, the event rate is
considerably lower than that at the proposed maximum energy of LEP2.
Nevertheless it has been advocated that such a scan should be done and,
indeed, that it could offer the most precise determination of $M_W$.

The precision determination of $M_W$ (and also the $W$ boson width
$\Gamma_W$) by means of a threshold scan relies on an accurate theoretical
knowledge of the electroweak radiative corrections to the $e^+ e^-
\rightarrow W^+ W^-$ cross section. Among these radiative effects, special
attention must be
paid to the electromagnetic Coulomb interaction between the $W^\pm$ bosons,
which results in the largest loop corrections in the important threshold
region.
It has been known for a long time \cite{SS} that when oppositely charged
particles have low relative velocity $v \ll 1$ (in units of $c$) Coulomb
effects enhance the cross section by a factor, which, to leading order in
$\alpha/v$, is ($1 + \alpha\pi/v$), provided that the particles are stable.
Subsequently it was shown \cite{FK1,FK2} that the Coulomb effects may be
radically modified when the interacting particles are short-lived rather than
stable. A general prescription which shows how to account for instability
effects in the threshold production of heavy particles was presented in Ref.
\cite{FKM3}. For the particular case of instability effects in $e^+e^-
\rightarrow W^+W^-$ near threshold a prescription was explicitly given
in a non-relativistic
framework in Ref. \cite{FKM1}. The results were only formulated in the
non-relativistic regime and so it is desirable to present a formulation
which covers the whole energy region. Such an attempt has been made
in Ref. \cite{BBD}, but unfortunately this extension was not correct,
as we shall explain
in Section 4. Here we present a complete treatment of the effects of
$W$ instability on the cross section for $e^+e^- \rightarrow W^+W^-$.

In Section 2 we present a qualitative discussion of Coulomb effects in
$e^+e^- \rightarrow W^+W^-$. In fact, general arguments allow us to identify
the energy domain in which $W$ instability will radically dilute the Coulomb
enhancement. In Section 3 we explicitly calculate the Coulomb effects for
$e^+e^- \rightarrow W^+W^- \rightarrow f_1 \bar f_2 f_3 \bar f_4$. We first
give a full non-relativistic derivation appropriate to the threshold region
and then we extend the formulation to include instability effects at any
energy. We discuss the level of accuracy to which these effects are included.
At each stage we check that the $\Gamma_W \rightarrow 0$ limit of our results
reproduces the Coulomb enhancement factor for stable $W$ production.
Whenever possible we give physical insight into the effects of $W$
instability on $W^+ W^-$ production. Although it is more physical and
transparent to use non-relativistic perturbation theory to calculate the
Coulomb effects between the $W^\pm$ bosons, in Section 4 we present an
alternative derivation based on Feynman diagram techniques. Again we explain
how to allow for the effects of instability of $W$ bosons at all $e^+ e^-$
energies. In Section 5 we show how to
calculate the higher-order Coulomb effects in the production of heavy
unstable particles \cite{FK2,FKM2,FKK}, and finally in Section 6 we present
our conclusions.

\newpage
\noindent{\large{\bf 2. Overview of Coulomb effects in $W^+W^-$ production}}
\vspace*{.5cm}

In this paper we wish to study the effects of the Coulomb interaction
between the $W^+$ and $W^-$ bosons in the process
\begin{equation}
e^+e^- \rightarrow W^+W^- \rightarrow (f_1\bar{f}_2)(f_3\bar{f}_4)
\label{eq:A}
\end{equation}
If the $W$ bosons were stable the effect of the Coulomb interaction on the
cross section has been known for a long time \cite{SS}. We summarize this
result in the subsection below, which allows us to establish notation. Then,
in the following subsection we discuss the modifications which arise from
the instability of the $W$ bosons. Using general arguments we show that
the modifications are particularly significant near the $W^+ W^-$ production
threshold, but become negligible for c.m. energies which satisfy
$\sqrt{s} - 2M_W \gg \Gamma_W$.

We shall not discuss the effects of initial state radiation.
These are very important, because of the logarithmic enhancements, but they can
be easily incorporated using standard structure function techniques \cite{KUF}
which are by now quite routine (see, for example, ref. \cite{FKM1}).   They do
not influence the qualitative features of the phenomena discussed here.  The
results presented below can equally well be directly applied to the process
$\gamma\gamma \rightarrow W^+W^-$.

\vspace*{.75cm}
\noindent {\bf (a) $e^+e^- \rightarrow W^+W^-$: assuming stable $W^{\pm}$
bosons}
\vspace*{.5cm}

For future reference we first study the hypothetical case in which the
$W^{\pm}$
bosons are assumed to be stable, that is we switch off the interaction
responsible for their decays.  In this $\Gamma_W \rightarrow 0$ limit the
inclusive $e^+e^- \rightarrow W^+W^-$ cross section can be written symbolically
in the form
\begin{equation}
\sigma(s) = \sigma_0(s,M^2_W,M^2_W)(1 + \delta(R,C))
\label{eq:B}
\end{equation}
where $\sigma_0$ is the cross section at c.m.\ energy $\sqrt{s}$ in the Born
approximation and $\delta(R,C)$ represents the radiative corrections.  The
suggestive form of the notation $\delta(R,C)$ implies that the Coulomb
corrections $C$ can be separated from the remaining radiative corrections $R$.
In fact the separation can only be done uniquely near
threshold where the two $W$'s are slowly moving in their c.m.\ frame
\cite{FKM3,FKM1}.  However it is just in the threshold region where the
off-shell and finite width effects are most important \cite{FK1,FK2,SP}.

For $W^+W^-$ S-wave production the separation of the exact (all-order)
Coulomb contribution and the first-order hard correction may be written
in the form\footnote{To the best of our knowledge, this type of factorized form
was first proposed in Ref. \cite{HB}; see Refs. \cite{FK2,FKS} for subsequent
discussions.}
\begin{equation}
1 + \delta(R,C) = |\psi(0)|^2 \left( 1 + \frac{\alpha}{\pi} \delta_H(s) \right)
,
\label{eq:C}
\end{equation}
with the Coulomb enhancement factor $|\psi(0)|^2$ separated from the
\lq\lq hard" or
short-distance one-loop electroweak corrections, which are denoted by
$\alpha\delta_H/\pi$.  The Coulomb factor, which was originally obtained by
Sommerfeld and Sakharov \cite{SS}, is given by
\begin{equation}
|\psi(0)|^2 = \frac{X}{1 - e^{-X}} = 1 + \frac{X}{2} + ...
\label{eq:D}
\end{equation}
with $X = 2\alpha\pi/v_0$, where $\psi(0)$ is the wave function, describing the
relative motion of the two $W$ bosons, evaluated at the origin.  The
S-wave configuration arises from the $\nu$ exchange diagram for $e^+e^-
\rightarrow W^+W^-$.  Here
\begin{equation}
v_0 = \frac{4p_0}{\sqrt{s}} = 2 \sqrt{1 - \frac{4M_W^2}{s}}
\label{eq:D1}
\end{equation}
is the relative velocity of the two $W$ bosons and $p_0$ is the magnitude
of the 3-momentum of each $W$ boson.  The subscript 0 is to indicate
that the bosons are on-mass-shell.  Later we will need to introduce the
off-shell velocity $v$ and momentum $p$. Instead of working in terms of the
c.m. energy $\sqrt{s}$ it is more convenient to introduce an energy
\begin{equation}
E = \frac{p_0^2}{M_W} = \frac{s - 4M_W^2}{4M_W}
\label{eq:F14}
\end{equation}
which coincides with the kinetic energy of the on-shell $W$ bosons in the
non-relativistic regime.

 From (\ref{eq:C}) and (\ref{eq:D}) we see that Coulomb
effects enhance the cross section by a factor which, to leading-order in
$\alpha/v$, is
\begin{equation} \left( 1 + \frac{\alpha}{v} \delta_C \right)
\label{eq:E}
\end{equation}
with $\delta_C = \pi$ and $v = v_0$ for {\it stable} $W$ bosons.
Therefore the short-distance correction in (\ref{eq:C}) is
\begin{equation}
\frac{\alpha}{\pi}\ \delta_H = \frac{\alpha}{\pi}\ \delta_1(s) -
\frac{\alpha\pi}{v_0}\ ,
\label{eq:F}
\end{equation}
where $\alpha\delta_1/\pi$ is the {\it full} one-loop electroweak correction to
$e^+e^- \rightarrow W^+W^-$.

It is interesting to note from (\ref{eq:D}) that at threshold the exact Coulomb
enhancement is twice the leading-order contribution.  This old result is
frequently overlooked in recent publications.

\vspace*{.75cm}
\noindent  {\bf (b) Coulomb effects in $e^+e^- \rightarrow W^+W^- \rightarrow
4f$: qualitative discussion}
\vspace*{.5cm}

The Coulomb correction is radically modified in the realistic case of unstable
$W$ bosons due to finite width and off-shell effects \cite{FKM1,FKM2}.  The
generic form of the cross section for $e^+e^- \rightarrow W^+W^- \rightarrow
4f$
is
\begin{eqnarray}
\sigma(s) = \int^s_0 ds_1 \rho(s_1) \int^{(\sqrt{s}-\sqrt{s_1})^2}_0
&&\!\!\!\!\!\!\!\!\! ds_2 \rho(s_2)
\sigma_0(s,s_1,s_2)(1 + \delta(R,\bar{C})) \, + \nonumber  \\
&& + \sum_n O\left( \alpha^n \frac{\Gamma_W}{M_W} \right)
\label{eq:G}
\end{eqnarray}
where $\sqrt{s_1}$ and $\sqrt{s_2}$ are the c.m.\ energies of the decay
products
of the $W$ bosons, and the Breit-Wigner factors
\begin{equation}
\rho(s_i) = \frac{B(W \rightarrow f\bar{f})}{\pi} \, \frac{\sqrt{s_i} \,
\Gamma_W(s_i)}{(s_i-M^2_W)^2 + s_i\Gamma^2_W(s_i)} .
\label{eq:H}
\end{equation}
The \lq\lq running" physical width $\Gamma_W(s_i) = \sqrt{s_i} \, \Gamma_W/M_W$
incorporates the radiative effects associated with the decays of the $W$
bosons.
In the limit $\Gamma_W \rightarrow 0$ and $B(W \rightarrow f\bar{f}) = 1$ we
see
(\ref{eq:G}) reduces to (\ref{eq:B}), providing, of course, that the modified
Coulomb effects, when averaged over the dominant regions of the $s_1,s_2$
integrations, reduce to the \lq\lq stable $W$ boson" result presented above.

The $n = 0$ term in the sum in (\ref{eq:G}) corresponds to the non-radiative
background contributions to $e^+e^- \rightarrow 4f$ (and their interferences
with the $e^+e^- \rightarrow W^+W^- \rightarrow 4f$ diagrams).  The $n \geq 1$
terms include the radiative interferences between the $W$ boson production and
decay stages induced by $n$ emitted quanta.  Note that the separation of the
terms of $O(\Gamma_W/M_W)$ is, in principle, gauge dependent.
As far as we are aware the complete analytical
calculation of the $O(\Gamma_W/M_W)$ interference terms has not been performed.

Apart from the appearance of the $O(\Gamma_W/M_W)$ terms, there are two
modifications of the \lq\lq stable $W$" formula (\ref{eq:B}), in going to the
realistic expression (\ref{eq:G}) in which we allow the $W$ bosons to decay.
First the obvious kinematic effect involving integrations over the Breit-Wigner
forms $\rho(s_i)$, and secondly the modification (symbolically denoted by $C
\rightarrow \bar{C}$) of the Coulomb interaction between the $W$ bosons, which
is our main concern.

It is straightforward to see at which $e^+e^-$ energies, $E$, the
modification of the Coulomb correction will be important.  The typical
interaction time between the $W$ bosons is $\tau_i \sim 1/(M_W v_0^2)$,
whereas the lifetime of the $W$ bosons is $\tau \sim
1/\Gamma_W$.  Therefore in the region $E \gg \Gamma_W$ we expect the Coulomb
effect to be unchanged by the instability of the $W$ bosons.  To be precise,
the modification could, at most, lead to a change of the cross section of
$O(\Gamma_W/(M_W v_0^2))$ in this region.

On the other hand in the threshold region $E \lapproxeq \Gamma_W$ the Coulomb
interaction time is comparable to, or even greater than, the $W$ lifetime.  We
therefore anticipate that the Coulomb correction will be considerably
suppressed
by the instability of the $W$ bosons.  In fact we can estimate the size of the
suppression of the original first-order Coulomb enhancement,
$(1 + \alpha\pi/v_0)$, as
follows.  We note from the expression for the cross section, (\ref{eq:G}), that
the interplay between the Breit-Wigner forms $\rho(s_i)$ and the phase space
factor in the Born cross section $\sigma_0$ (which is proportional to the c.m.\
momentum $p$ of a virtual $W$) suppresses contributions from the small momentum
region.  Indeed even in the threshold region we find
\begin{equation}
\langle p \rangle \gapproxeq \sqrt{M_W\Gamma_W} .
\label{eq:I}
\end{equation}
Thus, contrary to the stable $W$ case, we may say off-shell and finite width
effects mask the Coulomb singularity.  From (\ref{eq:I}) we see that the
expansion parameter of the Coulomb series is, at most,
\begin{equation}
\alpha \pi \sqrt{\frac{M_W}{\Gamma_W}} \sim 0.15 ,
\label{eq:J}
\end{equation}
rather than $\alpha\pi/v_0$ of the stable $W$ case.  Therefore higher-order
Coulomb corrections will be numerically small, although they can be calculated
exactly if necessary using the Green's function formalism of Refs.
\cite{FK2,FKM2,FKK}, see Section 5.

The first-order contribution to the radiative corrections $\delta(R,\bar{C})$
occurring in (\ref{eq:G}) can be written in the form
\begin{equation}
\delta^{(1)}(R,\bar{C}) = \frac{\alpha}{\pi}\ \delta_H(s)
+ \frac{\alpha}{v}\ \delta_C ,
\label{eq:K}
\end{equation}
which may be compared with $\delta(R,C)$ of eq. (\ref{eq:C}) for stable $W$
bosons.  Here, in the case of unstable $W$ bosons,
\begin{equation}
v = \frac{4p}{\sqrt{s}} = \frac{2}{s} \left [(s-s_1-s_2)^2 - 4s_1s_2\right ]
^{\frac{1}{2}}
\label{eq:L}
\end{equation}
and, similarly, $\delta_C$ depends on $s_1,s_2$, as well as $s$.  As before,
$p$
is the c.m.\ momentum of a virtual $W$ boson and $v$ is the relative velocity
of
the two bosons.
The \lq\lq hard" correction $\delta_H(s)$ was defined for stable $W$ bosons
in (\ref{eq:F}) for $s \geq 4M_W^2$. However in the unstable case
we can have values of $s$ below threshold.
As was discussed in Ref. \cite{FKM1} we can safely assume the threshold
value of $\delta_H$ in this region.

According to the above discussion, we do not expect the Coulomb
correction $\alpha\pi/v_0$ to be modified when $E \gg \Gamma_W$.  Indeed
in Section 3(c) we will show that, after
averaging over the dominant regions of the $s_1,s_2$ integrations in
(\ref{eq:G}), $\langle \delta_C\rangle \approx \pi$.  It
is easy to see from (\ref{eq:D1}) and (\ref{eq:L}) that
\begin{equation}
v = v_0 \left( 1 + O\left( \frac{\Gamma_W}{E} \right) \right)
\label{eq:M}
\end{equation}
for $E \gg \Gamma_W$, once we note that the dominant
$s_1$, $s_2$ integration regions are $|\sqrt{s_i}-M_W| \lapproxeq \Gamma_W$.
As a result the Coulomb correction $\alpha\pi/v_0$ for stable $WW$ production
is changed at most, as a result of instability, only by effects of relative
order $\Gamma_W/M_W v_0^2$ in the energy regime $E \gg \Gamma_W$.

\vspace*{.75cm}
\noindent{\large{\bf 3. Quantitative study of Coulomb effects}}
\vspace*{.5cm}

We start by calculating the Coulomb correction in the threshold region of
$W^+ W^-$ production since this is where the problem is well-defined and,
moreover, where the effects of instability of the $W$ bosons are most
important. We discuss the relativistic region at the end of the section.
For unstable particles we cannot restrict ourselves to on-shell
formula, and so we need to consider Green's functions rather than matrix
elements.

\vspace*{.75cm}
\noindent{\bf (a) Coulomb effects in the non-relativistic region}
\vspace*{.5cm}

The Green's function with two $W$ boson external legs is dependent on the two
off-shell variables, $s_1$, $s_2$, as well as on $s$. However in the
non-relativistic case we can reduce the problem to the evaluation of a
one-particle Green's function in an external field which depends only on one
off-mass-shell variable, $p^2 \neq M_W E$. As a consequence the Coulomb
factor in the matrix element for stable bosons,
$\psi (0)^*$, which depends just
on $E$, is replaced by $f({\bf p}, E)$ defined by \cite{FK2}
\begin{equation}
f({\bf p}, E)\ =\ <{\bf p} \vert (\widehat{H} - E - i\Gamma_W)^{-1} \vert
{\bf r} = 0> \left ({p^2\over M_W} - E - i\Gamma_W\right )
\label{eq:A3}
\end{equation}
where $\vert {\bf p}>$ is a $WW$ state of definite momentum ${\bf p} =
{\bf p} (W^+) = -{\bf p} (W^-)$, and $\vert{\bf r}>$ is a
state of definite relative position. The first factor on the right-hand
side of (\ref{eq:A3}) is the Fourier transform of the non-relativistic Green's
function $G_{E+i\Gamma_W} ({\bf r}^\prime,{\bf r})$ which decribes
the propagation of a $W^+ W^-$ pair created at relative distance ${\bf r}
= 0$. The second factor ensures that $f = 1$ in the absence of the Coulomb
interaction. The Hamiltonian in (\ref{eq:A3}) $\widehat{H} = \widehat{H}_0 +
\widehat{V}$
where
\begin{equation}
\widehat{H}_0 = {\widehat{{\bf p}}^2\over M_W}\,\,\,\, {\rm and} \,\,\,\,
\widehat{V} = -\frac{\hat\alpha}{r}.
\label{eq:B3}
\end{equation}

Before we proceed to evaluate (\ref{eq:A3}), it is informative to check that it
reduces to the stable, on-mass-shell result. To do this we use the
Lipmann-Schwinger equations for incoming $(\vert{\bf p}_+>)$ and outgoing
$(\vert{\bf p}_->)$ states
\begin{equation}
\vert{\bf p}_\pm> = \vert{\bf p}> + \left [ \left ( {{\bf p}^2\over M_W}
- \widehat{H}_0 \pm i \delta \right )^{-1} \widehat{V} \vert {\bf p}_\pm >
\right ] _{\textstyle \delta \rightarrow 0}.
\label{eq:C3}
\end{equation}
If we now express $\vert{\bf p}>$ in terms of $\vert{\bf p}_->$
then we can readily show from (\ref{eq:A3})
that the value of $f$ for on-shell, stable $W$ bosons is
\begin{equation}
\begin{array}[t]{cl}
\lim \\ {\scriptstyle \Gamma_W\rightarrow 0}
\end{array}
f ({\bf p}, E = {\bf p}^2/M_W) \ =\ <{\bf p}_- \vert {\bf r} = 0>
\ =\ \psi^-_{\bf p} (0)^*.
\label{eq:D3}
\end{equation}
Thus, since $\vert\psi (0)\vert^2 \equiv \vert \psi^\pm_{\bf p} (0) \vert^2$,
we recover the Coulomb enhancement factor for stable $W$ bosons\footnote{The
Coulomb interaction between incoming particles enhances the matrix element
by a factor $\psi^+_{\bf p} (0)$, while an interaction between outgoing
particles gives a factor $\psi^-_{\bf p} (0)^*$.}.

To calculate the Coulomb modification to the cross section, as defined in
(\ref{eq:E}), we expand $f({\bf p}, E)$ in terms of $V(r)$. From (\ref{eq:A3})
we obtain
\begin{eqnarray}
1 + {\alpha M_W\over 2p} \delta_C & \approx &\vert f({\bf p}, E) \vert^2
\nonumber \\
& = & 1 - 2{\rm Re} \int d^3 r e^{-i{\bf p}.{\bf r}} V(r) G^{(0)}_{E+i\Gamma_W}
({\bf r}, 0) + O(V^2)
\label{eq:E3}
\end{eqnarray}
where $G^{(0)}$ is the free-particle Green's function
\begin{eqnarray}
G^{(0)}_{E+i\Gamma_W} ({\bf r}, 0) & = & <{\bf r} \vert \left ( \widehat{H}_0
- E - i\Gamma_W \right )^{-1} \vert 0 > \nonumber \\
& = & {M_W\over 4\pi r} {\rm exp} (-\kappa r)
\label{eq:F3}
\end{eqnarray}
with
\begin{equation}
\kappa = \sqrt{M_W (-E - i\Gamma_W)} \equiv p_1 - ip_2.
\label{eq:G3}
\end{equation}
Solving for the real and imaginary parts, we have
\begin{equation}
p_{1,2} = \left [\ {\textstyle\frac{1}{2}} M_W \left ( \sqrt{E^2 + \Gamma^2_W}
\mp E\right ) \right ] ^{1\over 2},
\label{eq:H3}
\end{equation}
with $E$ given by (\ref{eq:F14}). We insert the Green's function
(\ref{eq:F3}) into (\ref{eq:E3}), and perform
the angular integration. We obtain
\begin{equation}
\alpha\delta_C = -2 \int^\infty_0 d r V(r) e^{-p_1 r} \lbrace \sin (p + p_2) r
+ \sin (p - p_2) r \rbrace
\label{eq:I3}
\end{equation}
where, at this stage, we have left $V(r)$ arbitrary so that
we will be better able to draw attention to the specific properties of the
Coulomb potential $V(r) = -\alpha/r$.

\vspace*{.75cm}
\noindent{\bf (b) Physical interpretation}
\vspace*{.5cm}

Before we present the analytical result of the integration in (\ref{eq:I3}),
it is informative to discuss some interesting features of $\delta_C$. Indeed
the interpretation of result (\ref{eq:I3}) for the Coulomb correction
$\delta_C$ is subtle and needs careful explanation. We begin by assuming
that the $W$ bosons are stable, $\Gamma_W = 0$ and $E > 0$.
Then from (\ref{eq:H3}) we have
\begin{equation}
p_1 = 0,\ \ \ \ p_2 = p_0 \equiv \sqrt{M_W E}.
\label{eq:J3}
\end{equation}
In this case (\ref{eq:I3}) can be readily evaluated using
\begin{equation}
\frac{2}{\pi} \int^\infty_0 \frac{dx}{x} \sin ax = {\rm sgn}\, a \equiv
\Biggl\lbrace
\begin{array}[c]{c}
\ 1\ {\rm for}\ a > 0 \\
\ 0\ {\rm for}\ a = 0 \\
-1\  {\rm for}\ a < 0,
\end{array}
\label{eq:K3}
\end{equation}
which yields
\begin{equation}
\delta_C = \pi \lbrace 1 + {\rm sgn} (p - p_0) \rbrace.
\label{eq:L3}
\end{equation}
We see that $\delta_C$ is a non-analytic function of the virtuality
$(p - p_0)$, with a {\it discontinuity} at the mass shell value
$p = p_0$.

The non-analytic behaviour of $\delta_C$ is a consequence of the
long-range nature of the Coulomb force, as we can verify by truncating the
potential so that $V(r) = 0$ for $r > R_0$, with $p_0 R_0 \gg 1$. For
the truncated potential, we find that $\delta_C$ makes a {\it smooth}
transition from 0 to $2\pi$ and that most of the variation occurs while the
virtuality
$(p - p_0)$ covers the range $\sim -1/R_0$ to $\sim 1/R_0$. Note
that the dominant contribution to the integral (\ref{eq:K3}) comes from
values $x \lapproxeq 1/\vert a\vert$.

We conclude that {\it any} non-zero virtuality will drastically change
the on-shell value $\delta_C = \pi$. This is contrary to explicit claims
presented in Ref. \cite{BBD}.

Now let us study the effect of the finite width $\Gamma_W$ of the $W$ bosons.
 From (\ref{eq:I3}) we see that the width plays the role of a cut-off
on the potential at distances $R_0 \sim 1/p_1$, where $p_1 \sim \Gamma_W
\sqrt{M_W/E}$ or $\sqrt{M_W \Gamma_W}$ according to whether $E$ is greater
or less than $\Gamma_W$. Thus, from the above discussion, we anticipate
that the presence of the finite width will lead to a smooth transition
between $\delta_C = 0$ and $\delta_C = 2\pi$ which occurs dominantly in
the region
\begin{equation}
\vert p - p_2\vert \sim p_1.
\end{equation}
The non-zero $W$ width thus restores the analyticity of $\delta_C$ as a
function of the $W$ boson virtuality.

Armed with this understanding, we return to (\ref{eq:I3}) and carry out the
integration explicitly. We find
\begin{equation}
\delta_C = \pi - 2\ {\rm arctan}\ \left (
\frac{p_1^2 + p_2^2 - p^2}{2 p p_1} \right ) ,
\label{eq:M3}
\end{equation}
a result which was obtained in Ref. \cite{FKM1} by a different approach.
Formula (\ref{eq:M3}) embodies all the special features of $\delta_C$ that
we have discussed above.

\vspace*{.5cm}
\noindent{\bf (c) Coulomb effects for $E \gg \Gamma_W$}
\vspace*{.5cm}

In Section 2(b) we used physical arguments to show that the instability of
$W$ bosons would not change the Coulomb enhancement factor of the $e^+ e^-
\rightarrow W^+ W^-$ cross section for energies $E \gg \Gamma_W$. At
first sight formula (\ref{eq:M3}) seems to contradict this claim, because
$\delta_C$ does not equal $\pi$
for $E \gg \Gamma_W$. However to determine the possible change of
{\it cross section} we must integrate the Coulomb correction $\delta_C$
over the $W$ boson virtualities $s_1$, $s_2$ as in (\ref{eq:G})\footnote
{The off-shell behaviour of the Coulomb effects in the unintegrated
cross section would be interesting to
observe, but, in practice, there will be insufficient statistics for
such a study.}. In the
non-relativistic case this reduces to an integration over the single off-shell
variable $p^2 \approx (\sqrt{s} - \sqrt{s_1} - \sqrt{s_2}) M_W$ \cite{FK2}
\begin{equation}
\int^s_0 ds_1 \rho(s_1) \int^{(\sqrt{s} - \sqrt{s_1})^2}_0 ds_2 \rho(s_2)
\approx \int^\infty_0 dp^2 \frac{\Gamma_W}
{\pi M_W [ (p^2/M_W - E)^2 + \Gamma^2_W ]}.
\label{eq:F10}
\end{equation}

For $E \gg \Gamma_W$ the arctan modification of $\delta_C$ of (\ref{eq:M3})
is an {\it odd} function of the virtuality $(p^2 - M_W E)$ in the dominant
region of the $p^2$ integration, which is specified by
$\vert p^2 - M_W E\vert \lapproxeq M_W \Gamma_W$, and hence integrates to
zero. To be explicit, for $E \gg \Gamma_W$ we find
\begin{equation}
\delta_C \approx \pi - 2\ {\rm arctan}\ \left ( \frac{M_W E - p^2}{M_W
\Gamma_W}
\right )
\label{eq:F11}
\end{equation}
in the essential $p^2$ region, and the difference of $\delta_C$ from the
on-mass-shell value of $\pi$ averages to zero when integrated
over the Breit-Wigner form in (\ref{eq:F10}).

\newpage
\noindent{\bf (d) Closed cross-section formula in the non-relativistic limit}
\vspace*{.5cm}

In the non-relativistic limit it is possible to use a mathematical trick
to carry out the off-shell integration and so obtain a closed formula for the
Coulomb corrections to the cross section. We wish to perform the
off-shell integration of (\ref{eq:F10}) over $\vert f({\bf p}, E)\vert^2$
weighted by the phase space factor $p/M_W$ occurring in $\sigma_0$. We
notice that the normalization factor for $f({\bf p}, E)$ given in (\ref{eq:A3})
cancels the Breit-Wigner denominator so that
\begin{eqnarray}
I_C &\equiv &\int^\infty_0\frac{pdp^2 \Gamma_W \vert f({\bf p}, E)\vert^2}
{\pi M_W^2 [ ( p^2/M_W - E)^2 + \Gamma_W^2]} \nonumber \\
&=& \frac{4\pi}{M_W^2}\ \int\ \frac{d^3p}{(2\pi)^3}\ \Gamma_W {\bigg\vert}
\langle {\bf p} \vert\ (\widehat{H} - E - i\Gamma_W )^{-1} \vert {\bf r} = 0
\rangle
{\bigg\vert}^2 \nonumber \\
&=& \frac{4\pi}{M_W^2}\ \langle {\bf r} = 0 \vert \frac{1}{\widehat{H} - E +
i\Gamma_W} \ \Gamma_W\ \frac{1}{\widehat{H} - E - i\Gamma_W} \vert{\bf r}
= 0 \rangle\nonumber \\
&=& \frac{4\pi}{M_W^2}\ {\rm Im}\ G_{E + i\Gamma_W} (0,0)\nonumber \\
&=& \frac{p_2}{M_W} + \alpha \ {\rm arctan} \left ( \frac{p_2}{p_1}
\right ) + O(\alpha^2),
\label{eq:T1}
\end{eqnarray}
where $p_{1,2}$ are defined in (\ref{eq:G3}).
Here we have used the completeness relation over the $\vert {\bf p} \rangle$
states and the explicit expression for the Green's function which can be
found in Ref. \cite{FK1}.

Formula (\ref{eq:T1}) clearly demonstrates that the modification of the cross
section due to $W$ instability is small in the region $E \gg \Gamma_W$,
since in this region
\begin{equation}
p_2 \approx p_0,\ \ \ \ \ \ \ \ p_1 \approx (\Gamma_W/E) p_0.
\label{eq:T2}
\end{equation}
Thus the arctan approaches $\pi/2$ and hence (\ref{eq:T1}) becomes up to
accuracy $O(\Gamma_W/E)$

\begin{equation}
I_C \approx \frac{1}{2}\ v_0 \left ( 1 + \frac{\pi\alpha}{v_0} \right )
\label{eq:T3}
\end{equation}
which corresponds to the leading-order stable boson result.

\vspace*{.75cm}
\noindent{\bf (e) Instability effects in the relativistic region}
\vspace*{.5cm}

The Coulomb correction is not uniquely defined and gauge
independent in the relativistic region.
For the production of stable $W$ bosons the Coulomb correction could equally
well be taken as
\begin{equation}
\frac{\alpha \pi}{v_0}\ \ \ \ \ {\rm or}\ \ \ \ \ \alpha\pi \left ( \frac
{M_W}{2p_0} \right )
\label{eq:F12}
\end{equation}
for example. In the relativistic domain we should not, therefore, discuss the
influence of $W$ instability on the {\it Coulomb} correction but rather its
possible modification of the {\it total} first-order correction. However once
we have agreed to {\it define} the Coulomb expansion parameter $X$ of
(\ref{eq:D}) as $2\alpha\pi/v_0$ and the \lq\lq hard" scattering correction
$\delta_H$ of (\ref{eq:C}) by (\ref{eq:F}), we need only focus on the
modification of the $\alpha\pi/v_0$ term. According to (\ref{eq:F}) and
(\ref{eq:K}) the modification, at leading order, is defined by adding the
expression
\begin{equation}
\frac{\alpha}{v}\ \delta_C\ - \ \frac{\alpha}{v_0}\ \pi
\label{eq:F13}
\end{equation}
to the correction for stable $W$ production. This difference gives, after
the integration in (\ref{eq:G}), the
modification of the cross section due to the instability of the $W$ bosons.

We have calculated $\delta_C$ in the non-relativistic domain where the Coulomb
correction is well defined and found that for $E \gg \Gamma_W$ the
modification is, at most, of relative order $\Gamma_W/E$. Thus it is evident
that the modification of the cross section in the relativistic domain
$E \gapproxeq M_W$ will be of relative order $\alpha\Gamma_W/M_W$, at most,
which is beyond our accuracy.

It is easy to check that if $\delta_C$ in the relativistic domain is defined
by (\ref{eq:M3}) it will satisfy the above criteria, {\it provided}
that the relativistic expressions (\ref{eq:D1}) and (\ref{eq:L}) are
used for $v_0$ and $v$ respectively, with $E$ defined by (\ref{eq:F14}).

\vspace*{.75cm}
\noindent {\large{\bf 4. Alternative derivation of the Coulomb effects}}
\vspace*{.5cm}

The calculation of the Coulomb correction that was presented in Section 3 was
based on non-relativistic perturbation theory. We believe this approach
is the most physical and transparent. Moreover the Coulomb interaction
is only well defined in the non-relativistic domain and, as we have seen,
this is the region in which the effects of instability are important.
Nevertheless it is informative to present an alternative derivation based
on Feynman diagram techniques.

In the Feynman diagram approach we have to evaluate the integral
\begin{eqnarray}
I = -i \int \frac{d^4k}{(2\pi)^4} &&\!\!\!\!\!\!\!\!\frac{1}
{(k^2 + i\varepsilon) [(k - p_-)^2 - M^2_W + iM_W \Gamma_W]} \times
\nonumber \\
&&\ \ \ \ \ \ \ \ \ \ \ \ \times
\frac{1}{[(k + p_+)^2 - M^2_W + iM_W \Gamma_W]}
\label{eq:Z1}
\end{eqnarray}
which corresponds to the loop diagram in which a photon of 4-momentum $k$
is exchanged between the two outgoing $W$ bosons of 4-momenta $p_\pm$, that is
$s_1 = p^2_+$ and $ s_2 = p^2_-$. We use Feynman parameter techniques to
reduce this integral to the form
\begin{eqnarray}
&I =& \frac{-1}{8\pi^2 s} \int^1_{-1} \frac{dx}{(x + (s_1 - s_2)/s)^2 -
\frac{1}{4} v^2} \times \nonumber \\
&&\times \log \left ( \frac{2\kappa^2 + \frac{1}{2} s x^2}
{2M_W^2 - s_1 - s_2 - 2iM_W\Gamma_W - (s_1 - s_2) x} \right )
\label{eq:Z2}
\end{eqnarray}
where $\kappa$ is defined in (\ref{eq:G3}) and $v$ is given by (\ref{eq:L}).
For arbitrary values of $v$ the result of the integration (\ref{eq:Z2})
cannot be expressed in terms of elementary functions, but instead involves
a sum of Spence functions (see, e.g. \cite{BBD}).
However we do not require the general expression for $I$
since it contains, besides the Coulomb effect, other contributions which
involve
infrared singularities etc. The extraction of the Coulomb part can only be done
unambiguously in the non-relativistic limit, and then only with an accuracy
up to terms of relative order $v$. Now it is clear from (\ref{eq:Z2}) that we
can extend the region of integration from $(-1, 1)$ to $(-\infty, \infty)$
without changing the result in the non-relativistic limit, provided that
$(s_1 - s_2)/s$ is small enough to preserve the $1/v$ Coulomb singularity.
Then the integral can be readily evaluated by making use of the analytic
properties of the integrand. First we note that the zeros of the denominator
in (\ref{eq:Z2}) are cancelled by the logarithm, since its argument becomes
unity at these points. The only singularities of the integrand of
(\ref{eq:Z2}) which remain are the branch points at $x = \pm 2i\kappa/\sqrt{s}$
and at
\begin{equation}
x = x_0 \equiv
\frac{2M_W^2 - s_1 - s_2 - 2iM_M\Gamma_W}{s_1 - s_2}.
\label{eq:Z3}
\end{equation}
We see that the position of $x_0$ depends on the sign of $s_1 - s_2$. If
$s_1 - s_2$ is positive (negative) then $x_0$ lies in the lower (upper) half
plane. This means that the $x_0$ branch point moves across the path of
integration and that our approximation (of extending the range of integration)
has destroyed the original analyticity of $I$ as a function of
$(s_1 - s_2)$. We shall see later some interesting consequences of this
observation.

We can now readily evaluate the integral $I$ of (\ref{eq:Z2}). If
$(s_1 - s_2) > 0$ we deform the contour of integration around the
cut starting at the branch point $x = 2i\kappa/\sqrt{s}$, and for
$(s_1 - s_2) < 0$ we wrap the contour around the cut starting from
$x = -2i\kappa/\sqrt{s}$. The result is
\begin{equation}
I = \frac{1}{4\pi i v s} \log \left ( \frac{i \kappa + {\textstyle\frac{1}{2}}
\sqrt{s} \Delta + p}{i \kappa + {\textstyle\frac{1}{2}} \sqrt{s} \Delta - p}
\right )
\label{eq:Z4}
\end{equation}
where $p$ is given by (\ref{eq:L}) and
\begin{equation}
\Delta \equiv \vert s_1 - s_2 \vert/s.
\label{eq:Z5}
\end{equation}
Since the first-order correction to the matrix element is proportional to
$I$ we are only interested in the real part of $I$. In fact the
Coulomb correction is
\begin{equation}
\delta_C^F = -8\pi v s\ {\rm Re}\ I = \pi - 2\ {\rm arctan}\ \left (
\frac{\vert i \kappa + {\textstyle\frac{1}{2}} \sqrt{s} \Delta \vert^2 - p^2}
{2 p p_1} \right )
\label{eq:Z6}
\end{equation}
where $\kappa \equiv p_1 -ip_2$ is given in (\ref{eq:G3}).
We use the superscript $F$ to distinguish the $\delta_C^F$ obtained from
Feynman diagram techniques from the $\delta_C$ which we calculated using
non-relativistic perturbation theory in (\ref{eq:M3}). Apart from the
occurrence of $\Delta$ in (\ref{eq:Z6}), the two results coincide.

A result identical to (\ref{eq:Z6}) was obtained by Bardin, Beenakker
and Denner
\cite{BBD}, also using the Feynman diagram approach. The two formulae for
$\delta_C^F$ can be seen to be the same if we note that their
$\beta_M = 2i\kappa/\sqrt{s}$. However the wrong conclusions were drawn in
Ref. \cite{BBD}, since $\Delta$ should be set to
zero, as we will show below.
An alert reader may have already guessed that this would be the case. If
we were to retain $\Delta$ then we would have a spurious singularity along
the line $s_1 = s_2$ (cf. (\ref{eq:Z5})); a singularity which was
introduced by the approximation used to evaluate the integral $I$ of
(\ref{eq:Z2}).

In summary we have presented two derivations of the Coulomb correction.
One based on
non-relativistic perturbation theory which gives $\delta_C$ of
(\ref{eq:M3}), see also Ref. \cite{FKM1}, and another based on Feynman
diagram techniques which gives $\delta_C^F$ of (\ref{eq:Z6}), see also
Ref. \cite{BBD}. The answers agree, except for the appearance of $\Delta$
in $\delta_C^F$. Both methods are, of course, equally correct; the difference
arises because of the approximations made in the derivation.

All these results have been obtained in the approximation that $v \ll 1$. Now
in the dominant range of the $s_1$, $s_2$ integrations in (\ref{eq:G}) we
see that
\begin{equation}
\Delta \equiv \frac{\vert s_1 - s_2\vert}{s} \sim \frac{\Gamma_W}{M_W}
\lapproxeq \langle v\rangle ^2,
\label{eq:Z7}
\end{equation}
recall eq.(\ref{eq:I}). So, on inspection of (\ref{eq:Z4}) and (\ref{eq:Z6})
we see that $\Delta$ may be neglected in comparison with $i\kappa/\sqrt{s}$
and $p/\sqrt{s}$. Therefore the two approaches, yielding $\delta_C$ and
$\delta_C^F$,
are entirely consistent in the non-relativistic region, as indeed they
must be. Thus whether or not we choose to retain $\Delta$ might appear to be
harmless.
This is true near threshold, but to extend the result away from the
$v_0 \ll 1$ region we must investigate the higher-order terms in $v$ to ensure
that we recover the formula for stable $W$ bosons in the limit that
$\Gamma_W \rightarrow 0$.

We rewrite the cross-section formula (\ref{eq:G}) in terms of the variables
\begin{equation}
x_i = \frac{s_i - M_W^2}{M_W\Gamma_W},\ \ \ \ \ {\rm with}\ \ \ i = 1,2.
\label{eq:Z8}
\end{equation}
If we then insert eq.(\ref{eq:K}) for $\delta(R,\bar C)$ we obtain
\begin{eqnarray}
\begin{array}[t]{cl}
\lim \\ {\scriptstyle \Gamma_W\rightarrow 0}
\end{array}
&\sigma(s)& = \sigma_0(s) \left ( 1 + \frac{\alpha}{\pi} \delta_1 (s)
\right ) +  \\
&&+ \sigma_0(s) \frac{\alpha}{v_0} \int^\infty_{-\infty} \frac{dx_1}
{\pi(x_1^2 + 1)} \int^\infty_{-\infty} \frac{dx_2}{\pi(x_2^2 + 1)}
\Bigl ( \delta_C (\Gamma_W = 0) - \pi \Bigr ) \nonumber
\label{eq:Z9}
\end{eqnarray}
where by $\delta_C (\Gamma_W = 0)$ we mean either $\delta_C$ of (\ref{eq:M3})
or $\delta_C^F$ of (\ref{eq:Z6}) expressed in terms of $x_1$, $x_2$ so
that the integrals can be performed in the limit $\Gamma_W \rightarrow 0$.
To evaluate $\delta_C$ we use (\ref{eq:L}), (\ref{eq:H3}) and (\ref{eq:F14})
and find
\begin{equation}
\Bigl ( \delta_C (\Gamma_W = 0) - \pi \Bigr ) = -2\ {\rm arctan}\
\left ( \frac{x_1 + x_2}{2} \right ).
\label{eq:Z10}
\end{equation}
The integration of this term gives zero and we recover the stable $W$ boson
result. Recall that $\alpha \delta_1/\pi$ is the {\it full} one-loop
correction, see (\ref{eq:F}). On the other hand for $\delta_C^F$ we find
\begin{equation}
\Bigl ( \delta_C^F (\Gamma_W = 0) - \pi \Bigr ) = -2\ {\rm arctan}\
\left ( \frac{x_1 + x_2}{2} + \frac{v_0}{2} \frac{\vert x_1 - x_2 \vert}{2}
\right )
\end{equation}
and the $\Gamma_W \rightarrow 0$ limit no longer reproduces the stable $W$
boson formula, as it should. In fact it is easy to see that in this case the
result gives
a value that is smaller than the Coulomb correction $\alpha\pi/v_0$ for
stable $W$ bosons. For $v_0 \ll 1$ it gives errors of relative order
$v_0$, but in the relativistic limit ${\textstyle\frac{1}{2}} v_0
\rightarrow 1$ it gives a Coulomb correction of ${\textstyle\frac{2}{3}}
(\alpha\pi/v_0)$ instead of $\alpha\pi/v_0$. This explains one of the
anomalies in Table 1 of Ref. \cite{BBD}.

The conclusion is that we must set $\Delta = 0$ and so formula
(\ref{eq:M3}), which was given previously in Ref. \cite{FKM1}, is correct
for all energies, provided that the appropriate relativistic definitions are
used for the kinematic variables\footnote{Unfortunately the authors of
Ref. \cite{BBD} used the formula in an incorrect way, which resulted in
a misleading comparison in their Table 1.}.

\vspace*{.75cm}
\noindent {\large{\bf 5. Higher-order Coulomb corrections}}
\vspace*{.5cm}

At the outset we should emphasize that the higher-order Coulomb corrections
to $e^+ e^- \rightarrow W^+ W^-$ are small and their exact (all-order)
calculation is beyond the needs of LEP2 today. The total contribution is less
than the existing uncertainties in the calculations of the $O(\alpha)$
\lq\lq standard" electroweak effects. Nevertheless, since Coulomb physics
(which is associated with large space-time intervals) is so different from the
other radiative effects, it merits study in its own right.

Section 3(a) already contains the appropriate formalism for calculating
the all-order Coulomb effect between unstable $W$ bosons. In analogy to
(\ref{eq:C}), the correction factor is
\begin{equation}
1 + \delta(R,\bar C) = \vert f({\bf p}, E) \vert^2
\left ( 1 + \frac{\alpha}{\pi} \delta_H (s) \right )
\label{eq:Y1}
\end{equation}
where $\delta_H(s)$ is defined by (\ref{eq:F}) and the non-relativistic
expression for $f({\bf p}, E)$ is given in (\ref{eq:A3}). Up to a factor
$(p^2/M_W - E - i\Gamma_W)$, $f({\bf p}, E)$ is the Fourier transform
of the non-relativistic Green's function; that is
\begin{equation}
f({\bf p}, E) = \left ( \frac{p^2}{M_W} - E - i\Gamma_W \right ) \int d^3 r\
e^{-i{\bf p}.{\bf r}} G_{E + i\Gamma_W} ({\bf r}, 0)
\label{eq:Y2}
\end{equation}
where
\begin{equation}
G_{E + i\Gamma_W} ({\bf r},{\bf r}^\prime) = \langle {\bf r} \vert
\left ( \widehat{H} - E - i\Gamma_W \right )^{-1} \vert {\bf r}^\prime \rangle.
\label{eq:Y3}
\end{equation}
This Fourier transform can be calculated with the help of the Meixner
representation \cite{MEI} of the Green's function $G_E ({\bf r}, 0)$.
It is found to be \cite{FKK}
\begin{equation}
f ({\bf p}, E) = 1 + 2\alpha M_W \kappa \int^1_0 dx
\frac{x^{-(\frac{1}{2}\alpha M_W/\kappa)}}{\kappa^2 (1+x)^2 +
p^2 (1 - x)^2}
\label{eq:Y4}
\end{equation}
where $\kappa$ is defined in (\ref{eq:G3}). It can be shown that the
integral in (\ref{eq:Y4}) is convergent for all real values of $E$,
provided that $\Gamma_W > 3\sqrt{3} M_W \alpha^2/32$ --- a condition easily
satisfied by the $W$ boson. Therefore the representation (\ref{eq:Y4}) is
applicable and well convergent for all real values of $E$, both below
and above the $WW$ threshold. The integrand in (\ref{eq:Y4}) has no
singularities in the interval $0 < x < 1$ and so $f({\bf p}, E)$ can be
readily computed numerically. Moreover we may expand $f({\bf p}, E)$ as a
power series in $\alpha$. It is easy to check that the real part of the
leading term is $\alpha M_W \delta_C / 4p$ with $\delta_C$ given by
(\ref{eq:M3}).

The identification of the leading term is the key to the generalisation of
representation (\ref{eq:Y4}) to the relativistic case. It can be done
by making the replacement $M_W \rightarrow \frac{1}{2} \sqrt{s}$ and by
using the correct relativistic expression for $p$ (cf. (\ref{eq:L})),
together with $\kappa$ defined in (\ref{eq:G3}) and $E$ given by
(\ref{eq:F14}).

\vspace*{.75cm}
\noindent {\large{\bf 6. Conclusions}}
\vspace*{.5cm}

The process $e^+ e^- \rightarrow W^+ W^-$ is one of the most fundamental
reactions to be studied at LEP2. It provides a unique opportunity to
probe the heart of the Standard Model, particularly if a precise
measurement of the mass $M_W$ of the $W$ boson can be obtained. For
the measurement of $M_W$ it is necessary to have an accurate
theoretical knowledge of the cross section, especially in the region of the
$WW$ threshold.

Here we calculate the corrections to the cross section for $e^+ e^-
\rightarrow W^+ W^-$ which arise from the instability of the produced
$W$ bosons. Although our result applies at all energies it is useful to
concentrate on the modification which occurs in the important
$W^+ W^-$ threshold region.
For stable $W$ bosons we may write the cross section in the symbolic form
\begin{eqnarray}
\sigma &\sim& f_{\rm ISR}\ \ f_{\rm EW}\ \ \vert \psi(0) \vert^2\ \ \sigma_0
\nonumber \\
&\sim& v_0^\lambda \left ( 1 + \frac{\alpha}{\pi} \delta_H \right )
\left ( 1 + \frac{\alpha\pi}{v_0} + ... \right ) v_0
\label{eq:X1}
\end{eqnarray}
where $v_0$ is the relative velocity of the $W$ bosons and $\sigma_0$ is
the Born cross section. Since initial state radiation (ISR) can be
included in a straightforward way we have regarded it as an inessential
complication and neglected it in our study. For completeness we show
its threshold behaviour in (\ref{eq:X1}), where $\lambda \approx 4\alpha
\log (s/m_e^2) / \pi$. The \lq\lq hard" or short-distance electroweak (EW)
corrections are also noted in (\ref{eq:X1}) for completeness. However
the threshold behaviour is dominated by the Coulomb enhancement factor
and the $v_0$ phase space factor in $\sigma_0$.

Clearly in order to precisely measure $M_W$ by an energy scan of the
cross section in the
threshold region it is crucial to calculate the modification of the Coulomb
effect arising from the instability of the $W$ bosons. It is easy to see that
the major modification will, in fact, occur in the threshold region.
Essentially
what happens is that the instability of the $W$ bosons smooths out the
Coulomb singularity on account of the intrinsic uncertainty in their
relative velocity $\sim \sqrt{\Gamma_W / M_W}$. We quantify the modification
due to instability in eqs. (\ref{eq:G}), (\ref{eq:K}) and ({\ref{eq:M3}).

The modification is very important for energies $E \lapproxeq \Gamma_W$,
but fades away as $E$ increases so that for $\Gamma_W \ll E \ll M_W$
the effect is of $O (\Gamma_W / E)$ at most, and for $E \gapproxeq M_W$
of $O (\Gamma_W / M_W)$ at most, where the energy $E$ is defined in
(\ref{eq:F14}). The formula that we present for the modification of the
cross section due to the instability of the $W$ bosons is unambiguous for
all energies\footnote{Clearly we must use the relativistic expressions
for the kinematic variables, that is (\ref{eq:L}) and (\ref{eq:H3}) with
$E$ defined as in (\ref{eq:F14}). We have shown in Section 4 that the variable
$\Delta$ introduced in Ref. \cite{BBD} must be set to zero. $\Delta$ is an
artefact of the approximation used to calculate the Coulomb effects and
introduces a spurious singularity, but, more important, it leads to an
increasingly incorrect result as the energy increases away from the
threshold region.}, despite the fact that the Coulomb
interaction can only be uniquely defined for $E \ll M_W$. Using the formulae
(\ref{eq:G}), (\ref{eq:K}) and ({\ref{eq:M3}) it is straightforward to
allow for the important $W$ boson instability effects in an experimental
study of $e^+ e^- \rightarrow W^+ W^-$ in the $WW$ threshold region.

\vspace*{.75cm}
\noindent{\large{\bf Acknowledgements}}
\vspace*{.5cm}

We thank W.J.\ Stirling for valuable discussions. V.S. Fadin thanks the Centre
for Particle Theory and Grey College of the University of Durham for
hospitality. The financial support of the UK Particle Physics and Astronomy
Research Council is gratefully acknowledged.

\end{document}